\documentclass[11pt]{article}
\PassOptionsToPackage{table}{xcolor}
\usepackage[preprint]{acl}
\usepackage{times}
\usepackage{latexsym}
\usepackage[T1]{fontenc}
\usepackage[utf8]{inputenc}
\usepackage{microtype}
\usepackage{inconsolata}
\usepackage{graphicx}
\usepackage{placeins}
\makeatletter
\let\swe@normaladdtocurcol\@addtocurcol
\let\swe@normalcombinedblfloats\@combinedblfloats
\makeatother
\usepackage{stfloats}
\makeatletter
\let\swe@appendixaddtocurcol\@addtocurcol
\let\swe@appendixcombinedblfloats\@combinedblfloats
\let\@addtocurcol\swe@normaladdtocurcol
\let\@combinedblfloats\swe@normalcombinedblfloats
\newcommand{\sweappendixfloats}{%
  \let\@addtocurcol\swe@appendixaddtocurcol
  \let\@combinedblfloats\swe@appendixcombinedblfloats
}
\makeatother
\usepackage{amsmath,amssymb,booktabs,multirow}
\usepackage{array,tabularx}
\usepackage[most]{tcolorbox}
\usepackage{enumitem}
\newcolumntype{Y}{>{\centering\arraybackslash}X}
\definecolor{sweviolet}{HTML}{817BA6}
\definecolor{sweteal}{HTML}{5F9E9A}
\definecolor{swepanel}{HTML}{F6F7FA}
\definecolor{sweproprietaryrow}{HTML}{F0EDF5}
\definecolor{sweopenrow}{HTML}{E8F2F0}
\definecolor{swescaffoldrow}{HTML}{EAF0F5}
\definecolor{swezebra}{HTML}{F7F8FA}
\newtcolorbox{swepromptbox}[2][]{
  enhanced,
  colback=swepanel,
  colframe=black!72,
  colbacktitle=black!78,
  coltitle=white,
  boxrule=0.45pt,
  arc=1.2pt,
  left=5pt,
  right=5pt,
  top=5pt,
  bottom=5pt,
  fonttitle=\bfseries\small,
  fontupper=\small,
  title={#2},
  #1
}
\newtcolorbox{swecasebox}[2][]{
  enhanced,
  colback=swepanel,
  colframe=black!75,
  colbacktitle=sweviolet,
  coltitle=white,
  boxrule=0.55pt,
  arc=1.5pt,
  left=6pt,
  right=6pt,
  top=6pt,
  bottom=6pt,
  fonttitle=\bfseries\small,
  fontupper=\small,
  title={#2},
  #1
}
\newenvironment{sweappendixblock}[1]{%
  \noindent\begin{minipage}[t]{#1}%
  \renewenvironment{table}[1][]{\captionsetup{type=table}}{\par}%
  \renewenvironment{table*}[1][]{\captionsetup{type=table}}{\par}%
  \renewenvironment{figure*}[1][]{\captionsetup{type=figure}}{\par}%
}{\end{minipage}}
\newcommand{\method}{\mbox{SWE-Nexus}}
\title{SWE-Nexus: Automated Composition of Software Engineering Tasks\\from Test Execution}
\author{
\textbf{Songwei Yu}$^{1,\star}$,
\textbf{Ziqi Miao}$^{2,\star}$,
\textbf{Jingxuan Xu}$^{4,\star}$,
\textbf{Junhao Fan}$^{3}$,
\textbf{Haoyang Huang}$^{4}$ \\
\textbf{Yuan Xiong}$^{2}$,
\textbf{Zihang Xu}$^{2}$,
\textbf{Weihao Li}$^{4}$,
\textbf{Wenqiang Zhu}$^{4}$,
\textbf{Yuechen Luo}$^{3}$ \\
\textbf{Xinshun Feng}$^{2}$,
\textbf{Yanan Wu}$^{4}$,
\textbf{Ken Deng}$^{4,\dagger}$ \\
$^1$ National University of Singapore, Singapore \\
$^2$ Shanghai Artificial Intelligence Laboratory, Shanghai, China \\
$^3$ Tsinghua University, Beijing, China \\
$^4$ Independent Researchers \\
\texttt{mzq19375241@gmail.com}
}

\hypersetup{
  pdftitle={SWE-Nexus: Automated Composition of Software Engineering Tasks from Test Execution},
  pdfauthor={Songwei Yu, Ziqi Miao, Jingxuan Xu, Junhao Fan, Haoyang Huang, Yuan Xiong, Zihang Xu, Weihao Li, Wenqiang Zhu, Yuechen Luo, Xinshun Feng, Yanan Wu, Ken Deng}
}

\begin{document}
\maketitle
\begingroup
\renewcommand{\thefootnote}{}
\makeatletter
\renewcommand{\@makefntext}[1]{\noindent #1}
\makeatother
\footnotetext{%
$^{\star}$ Equal contribution.\\
$^{\dagger}$ Corresponding author.}
\endgroup
\begin{abstract}
Historical code changes provide convenient task boundaries, but do not identify which production functions selected tests exercise or support composing validated behaviors beyond a single change. We present \method{}, a fully automated pipeline that constructs repository-level implementation tasks from test execution. Four construction agents select tests, adapt tracing and evaluation protocols to each repository, and generate behavioral specifications. For each selected test file, tracing records an observed execution footprint of production function bodies. A scope is retained only if all expected scoring points pass in the original repository and masking the traced bodies causes at least one failure or error. Compatible scopes are composed deterministically, with shared functions counted once and every constituent test file revalidated under the complete combined mask. The benchmark contains 600 tasks from 105 repositories across six programming languages, including 116 atomic and 484 composite tasks. We evaluate 12 models under three agent scaffolds. The best configuration resolves 37.8\% of tasks in one attempt, and scope-weighted scoring measures partial recovery. \method{} enables task construction from validated behavior beyond individual historical changes.

\end{abstract}
\section{Introduction}
\label{sec:introduction}

\begin{figure}[!t]
  \centering
  \includegraphics[width=\columnwidth]{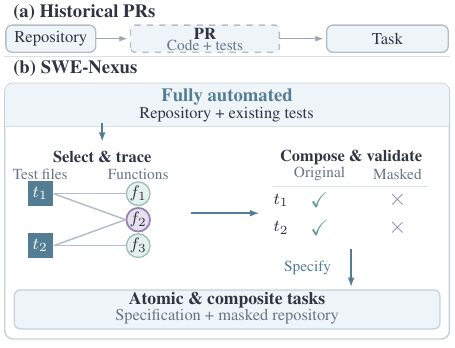}
  \caption{Task construction from historical PRs (a) and test execution (b). Edges show observed test--function associations. Deterministic composition counts shared functions once. Independent checks before and after composition require all expected points to pass originally~($\checkmark$) and at least one failure or error per file under the complete mask~($\times$).}
  \label{fig:intro-concept}
\end{figure}

Repository-level benchmarks commonly derive tasks from historical pull requests (PRs), combining issue descriptions, code changes, tests, and execution environments \citep{jimenez2024swebench,zan2025multiswebench,badertdinov2026swerebenchv2}. As agents solve more established instances, fixed collections become less informative about their capabilities \citep{deng2026swebenchpro}. Public implementations also raise concerns about training data contamination \citep{badertdinov2025swerebench,jain2025livecodebench}. Informative evaluation and execution-based training therefore require a continuing supply of challenging, verifiable tasks \citep{pan2025swegym,yang2025swesmith,jain2025r2egym,wei2026selfplayswerl}. Yet code and tests that appear in the same PR do not reveal which implementation each test exercises. For example, a test may exercise only some modified functions while calling unchanged helpers. Historical patch extraction can also include unrelated tests \citep{badertdinov2025swerebench}. Human review can improve task descriptions and test suites \citep{deng2026swebenchpro}, while execution tracing directly records the functions each test exercises \citep{rothermel1997regression}. A PR or commit captures one historical change, whereas a software requirement may span files and modules.

\begin{samepage}
\noindent These challenges motivate a central question:

\emph{Can we automatically construct diverse and challenging repository-level tasks beyond individual PR boundaries, with implementation scopes validated through test execution?}\par
\end{samepage}

We introduce \method{}, a fully automated, four-agent pipeline that constructs atomic and composite implementation tasks from test execution (Fig.~\ref{fig:intro-concept}). The \emph{Selector}, \emph{Trace}, \emph{Validation}, and \emph{Query Agents} handle test selection, tracing, protocol adaptation, and specification generation, respectively. The production function bodies recorded for each test file form its \emph{observed execution footprint}. An independent procedure retains a scope only when all expected scoring points pass in the original repository and masking the recorded bodies causes at least one point to fail or report an error. Each validated scope forms an atomic task. Compatible scopes from the same repository revision and environment are combined deterministically into composite tasks. Composition counts shared functions once and revalidates every constituent test file under the complete combined mask.

\begin{samepage}
The evaluated solver is independent of the construction agents and receives only the behavioral specification and masked repository. Selected scoring tests and construction metadata remain hidden. The benchmark contains 600 tasks from 105 repositories, balanced across six programming languages. We evaluate 12 models under three agent scaffolds using the resolved rate for complete recovery and the scope-weighted score for partial recovery. The best configuration resolves 37.8\% of tasks in one attempt. Under Claude Code, mean resolved rates decline from 55.2\% on Easy tasks to 7.4\% on Hard tasks. Within comparable configurations, resolved attempts generally have lower median counts of tool calls and modified files.\par
\end{samepage}

Our contributions are threefold:
\begin{itemize}
  \item \textbf{Automated construction.} We introduce a fully automated four-agent pipeline that discovers implementation scopes from test execution. Fixed procedures validate each scope and recheck every constituent test file after deterministic composition.
  \item \textbf{Compositional benchmark.} We construct 600 tasks across six languages: 116 atomic and 484 composite tasks spanning two to nine test files.
  \item \textbf{Evaluation.} We introduce scope-weighted scoring for partial recovery, evaluate 12 models with three agent scaffolds, and analyze recorded solver activity.
\end{itemize}

\section{Related Work}
\label{sec:related-work}

\paragraph{Repository-level evaluation.}
Defects4J and BugsInPy provide reproducible bugs and tests for controlled testing and debugging \citep{just2014defects4j,widyasari2020bugsinpy}. SWE-bench evaluates patches for GitHub issues using repository tests \citep{jimenez2024swebench}. Multi-SWE-bench broadens language coverage, while SWE-rebench V2 automates repository setup and task-quality assessment \citep{zan2025multiswebench,badertdinov2026swerebenchv2}. SWE-rebench collects fresh tasks for contamination-sensitive evaluation \citep{badertdinov2025swerebench}, and SWE-Bench Pro targets longer-horizon changes with explicit requirements and interfaces \citep{deng2026swebenchpro}.

\paragraph{Constructing executable tasks.}
SWE-smith retains model-generated, procedural, and historical modifications that break previously passing tests, then combines bugs within files or modules \citep{yang2025swesmith}. Self-Play SWE-RL trains through iterative bug injection and repair with test patches as specifications \citep{wei2026selfplayswerl}. R2E-Gym synthesizes tests and problem descriptions from commits \citep{jain2025r2egym}. FeatureBench constructs feature development tasks across multiple PRs by tracing Python tests \citep{zhou2026featurebench}.

\paragraph{Agent scaffolds and training.}
SWE-agent and OpenHands provide interfaces for inspecting, editing, and executing repository code, whereas Agentless uses a fixed localization, repair, and validation pipeline \citep{yang2024sweagent,wang2025openhands,xia2025agentless}. SWE-agent's interface ablations examine how search, editing, and execution feedback affect task resolution and repository navigation \citep{yang2024sweagent}. SWE-Gym and R2E-Gym train agents and verifiers in executable repository environments \citep{pan2025swegym,jain2025r2egym}, where test outcomes supply verifiable behavioral feedback \citep{le2022coderl}. SWE-RL learns from historical changes using patch-similarity rewards, while DeepSWE trains coding agents with test-based execution rewards \citep{wei2025swerl,luo2025deepswe}.

% Keep the two large construction figures on separate pages.
\setcounter{dbltopnumber}{1}

\begin{figure*}[t]
  \centering
  \includegraphics[width=\textwidth]{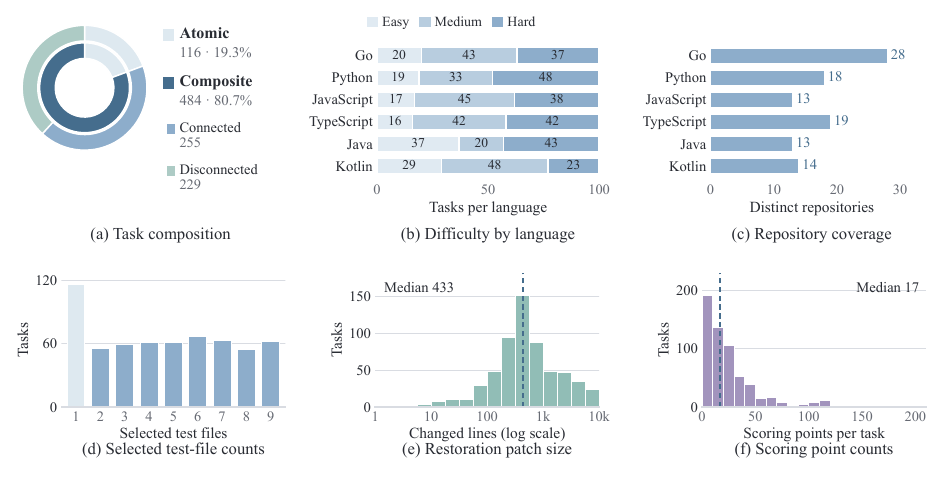}
  \caption{Statistics of the 600-task benchmark: (a) task composition and connectivity of composite tasks (connected or disconnected), (b) difficulty by language (100 tasks each), (c) repository coverage, (d) selected test-file counts, (e) changed lines in reference restoration patches (log scale), and (f) scoring points per task (10-point bins). Edges link test files that share observed production functions. Dashed lines mark medians.}
  \label{fig:dataset-statistics}
\end{figure*}

\section{Dataset Overview}
\label{sec:overview}

Each \method{} task asks a solver to recover repository functionality from a natural-language specification. The surrounding codebase remains available, with selected production function bodies masked and scoring tests withheld.

\subsection{Task Definition}
\label{sec:task-setting}

Let $R$ denote a repository at a fixed revision with an associated execution environment. A task is defined by selected test files $\mathcal{T}$ and a validated implementation scope $S$ of production functions whose bodies are masked. Let $R^{-}$ denote the masked repository. The solver receives a natural-language specification $q$ and a copy of $R^{-}$ with the selected test files withheld. It submits a patch $\Delta$ to production code for evaluation against those tests in a clean environment. Unselected repository tests may remain available during development.

The construction procedure retains the original implementations, selected tests, and fixed evaluation protocol. Restoring the original bodies gives a reference restoration patch for measuring task size. The tests and protocol define a fixed set of expected scoring points $\mathcal{P}$. A scoring point is an evaluation unit with an outcome that can be identified consistently, typically a test case or parameterized instance. When finer outcomes are unavailable, a selected test file serves as one scoring point. Individual assertions are not counted separately. Evaluation tests the specified behavior without requiring the original implementation.

\begin{figure*}[t]
  \centering
  \includegraphics[width=\textwidth]{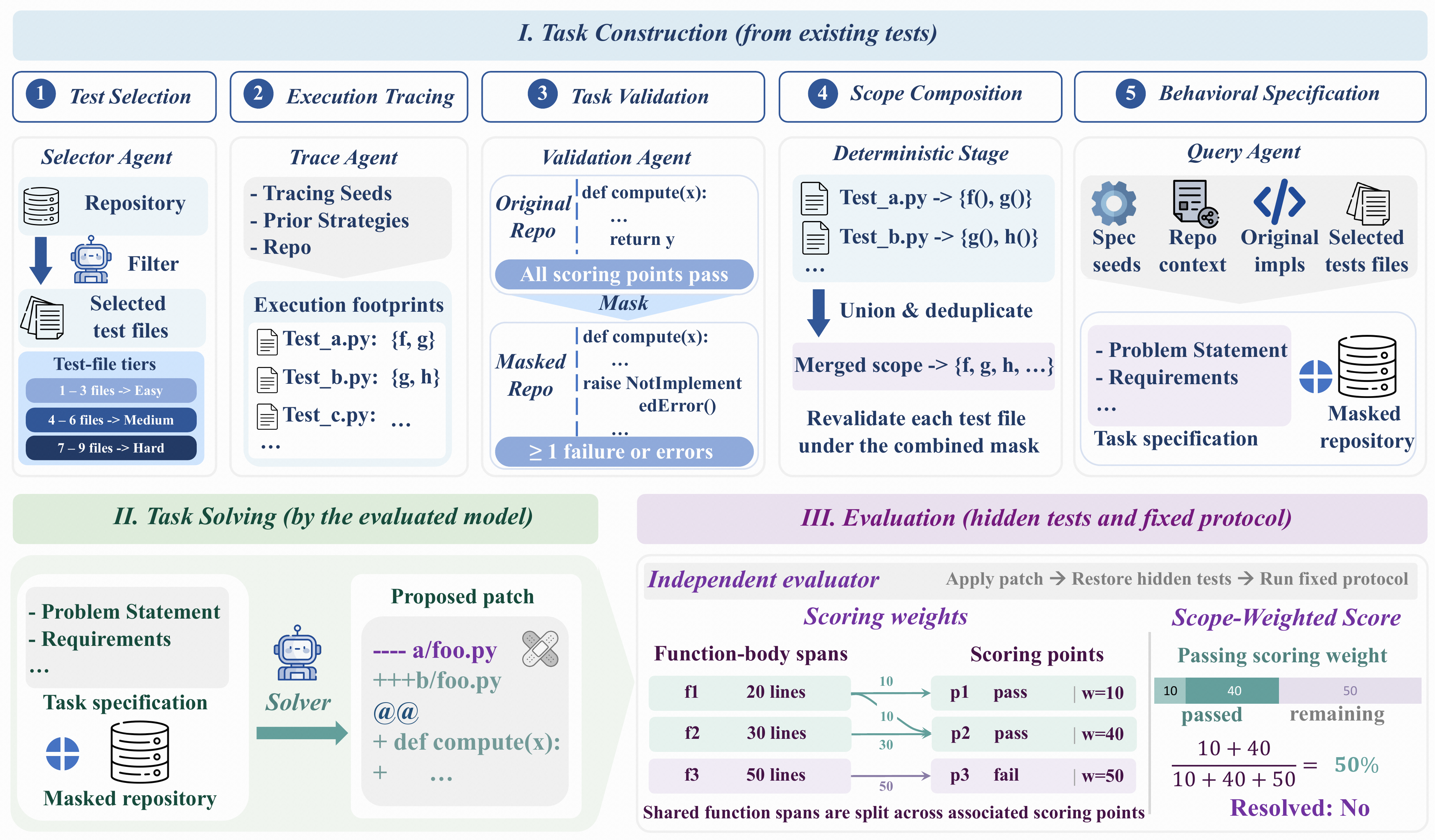}
  \caption{Construction and evaluation in \method{}. The Selector, Trace, Validation, and Query Agents select tests, trace execution, adapt protocols, and generate specifications. Independent checks validate scopes before and after composition. The solver receives the specification and masked repository, and the evaluator restores hidden tests and applies the fixed protocol. Passing $p_1$ and $p_2$ gives a scope-weighted score of 50\% and leaves the task unresolved.}
  \label{fig:method}
\end{figure*}

\subsection{Dataset Characteristics}
\label{sec:benchmark}

The benchmark contains 600 tasks from 105 repositories across six languages, with 100 tasks per language: 116 atomic tasks use one test file, and 484 composite tasks combine two to nine (Fig.~\ref{fig:dataset-statistics}(a)). Linking test files that share observed production functions yields one connected component (\emph{connected}) or multiple components (\emph{disconnected}). Behaviors span configuration parsing, application interfaces, and data serialization. One cloud-init task, for example, combines hosts-file operations with resolver configuration.

Tasks have a median of five selected test files and 35 masked production function bodies. Reference restoration patches modify a median of 10 production files and 433 lines, counting additions and deletions. The interquartile ranges are 2--7 selected test files, 242--1,046 changed lines, and 7--32 scoring points. The full range is 1--202 scoring points, totaling 15,975 task--point pairs across the release. Selected tests and their scoring points may recur across composite tasks.

\paragraph{Task scope.}
We assign Easy, Medium, and Hard tiers using selected test-file count (1--3, 4--6, 7--9) and reference restoration patch size (1--300, 301--999, $\geq 1{,}000$ changed lines). The final difficulty is the higher tier. Every language includes all three groups (Fig.~\ref{fig:dataset-statistics}(b) and Appendix~\ref{app:data}).

\section{Dataset Construction}
\label{sec:method}

\method{} links repository tests to the production code they exercise. For a fixed revision and execution environment, tracing identifies candidate scopes, and validation retains those whose masking disrupts expected behavior. These scopes form atomic tasks or are composed deterministically into composite tasks. Four construction agents select tests, adapt tracing and evaluation protocols, and generate specifications. Masking, verification, composition, and scoring follow fixed procedures (Fig.~\ref{fig:method}).

\subsection{Execution-Grounded Scope Validation}
\label{sec:tracing}
\label{sec:validation}

Test assertions specify expected behavior, but do not identify which production code to mask. The \emph{Selector Agent} chooses files from a prefiltered pool, prioritizing explicit behavioral assertions, stable execution, and focused functional scope. The \emph{Trace Agent} connects these tests to production code through runtime observation. It uses high-quality tracing examples, existing strategies, and repository instructions to adapt tracing to the test runner, process boundaries, and source layout. For each selected file $t$, it executes the tests and maps observed functions to editable production function bodies. The resulting set $S_t$ is the \emph{observed execution footprint} for $t$. We retain these test--function associations for composition and scoring.

To verify that masking the observed scope affects the tested behavior, we replace the bodies in $S_t$ with language-appropriate stubs, producing $R_t^{-}$ while preserving declarations, signatures, and surrounding code structure. The \emph{Validation Agent} adapts the execution plan and result parser to the repository and fixed environment. The resulting protocol specifies how to execute tests, parse their outputs, and assign each expected scoring point a pass, failure, or error status. An independent validation procedure runs this protocol on $R$ and $R_t^{-}$. Let $\mathcal{P}_t$ denote the expected scoring points for $t$, and let $o_X(p)$ denote the parsed outcome of point $p$ in repository state $X$. We retain a scope only if
\begin{equation}
\begin{aligned}
&\forall p\in\mathcal{P}_t,\quad o_R(p)=\mathrm{pass},\\
&\exists p\in\mathcal{P}_t:\quad o_{R_t^{-}}(p)\in\{\mathrm{fail},\mathrm{error}\}.
\end{aligned}
\label{eq:admission}
\end{equation}
The original run checks test validity, while the masked run checks whether removing the scope causes an observable failure or error. Each validated scope is retained with its evaluation protocol.

\subsection{Deterministic Scope Composition}
\label{sec:composition}

One validated scope forms an atomic task with a single selected test file. Combining scopes requires counting shared functions once and ensuring that every constituent test remains relevant. For compatible test files $\mathcal{T}$ from the same repository revision and execution environment, we take the union \mbox{$S=\bigcup_{t\in\mathcal{T}}S_t$}. Compatibility requires that the scopes can be masked and evaluated together. We deduplicate shared functions by their source identities and build the complete masking patch from the clean original source.

The independent validation procedure rechecks every constituent file under the complete combined mask using the fixed execution and parsing rules. For each file $t\in\mathcal{T}$, all expected scoring points must pass in the original repository, and at least one must fail or report an error in the masked repository. Every selected file must therefore test behavior affected by the final mask, including files that share functions. Validated composite tasks contain two to nine selected test files and require one implementation to satisfy their combined requirements. This composition stage is deterministic.

\subsection{Behavioral Specification}
\label{sec:task-description}

The natural-language specification describes the behavior to recover while the selected scoring tests remain hidden. The \emph{Query Agent} generates the specification $q$ from the original implementations, selected tests, and repository context. High-quality specification examples guide its structure and coverage. The \emph{Problem Statement} explains the target functionality, its purpose, and its role in the repository. The \emph{Requirements} describe observable inputs, outputs, boundary cases, defaults, error handling, and side effects. The task package pairs $q$ with the masked repository $R^{-}$ after removing the selected test files. Execution traces, test--function associations, expected outcomes, and scoring configurations remain hidden. The solver must recover the behavior from the specification and available repository context.

\subsection{Fixed Evaluation and Scoring}
\label{sec:specification}
\label{sec:metrics}

The solver is the model under evaluation and operates independently of the four construction agents. It receives the task package and submits a patch $\Delta$ to production code. In a clean environment, an independent evaluator applies $\Delta$ to $R^{-}$, restores the selected test files from a trusted copy, and runs the protocol established during task validation. Test execution, output parsing, and scoring criteria are fixed before solving and applied consistently across submissions. A task is resolved only when the evaluation run is valid, every expected scoring point passes, and the patch introduces no additional failures detected by the protocol. The \emph{resolved rate} is the proportion of tasks satisfying this criterion.

To measure partial recovery, the \emph{scope-weighted score} weights test outcomes by the source spans of associated production function bodies. Each function $f\in S$ has a recorded body span of $L_f$ source lines. The retained associations map each function to a nonempty set of expected scoring points $A(f)\subseteq\mathcal{P}$ from tests whose footprints contain that function. Let $b_p(\Delta)\in\{0,1\}$ indicate whether point $p$ passes for patch $\Delta$. We divide each function's span equally among its associated points and fix the resulting weights before solving:
\begin{equation}
\begin{aligned}
w_p&=\sum_{f\in S:\,p\in A(f)}\frac{L_f}{|A(f)|},\\
S_w(\Delta)&=\frac{\sum_{p\in\mathcal{P}}w_p b_p(\Delta)}{\sum_{p\in\mathcal{P}}w_p},
\end{aligned}
\label{eq:weighted}
\end{equation}
Each function contributes its span once, including functions shared by several tests, so \mbox{$\sum_p w_p=\sum_f L_f$}. Appendix~\ref{app:metrics} describes the coarser and unit-weight configurations used when finer attribution is unavailable. These configurations are also fixed before solving.

\setcounter{dbltopnumber}{2}
\begin{figure*}[!t]
  \centering
  \includegraphics[width=\textwidth]{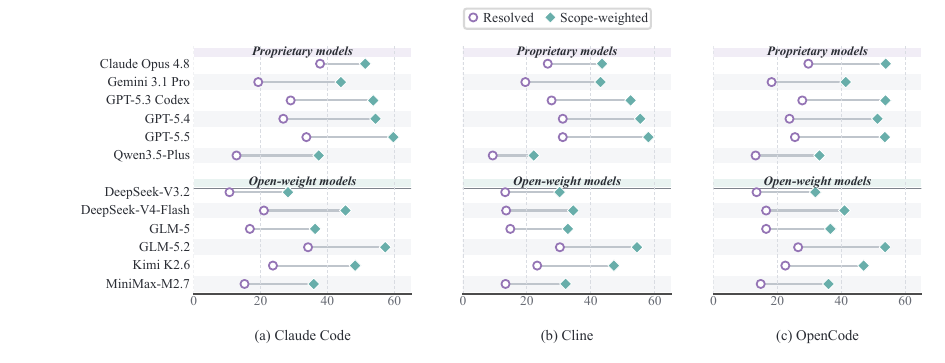}
  \caption{Overall P@1 performance across 12 models and three agent scaffolds. Segments connect each model's resolved rate (circle) and mean scope-weighted score (diamond). All panels share a 0--65\% scale. Appendix~\ref{app:evaluation-details} reports exact values.}
  \label{fig:overall-performance}
\end{figure*}

% Generated by scripts/plot_results.py; edit the source data or generator.
\begin{table*}[t]
\centering
\fontsize{7.35}{8.35}\selectfont
\renewcommand{\arraystretch}{1.06}
\setlength{\tabcolsep}{0.8pt}
\begin{tabularx}{\textwidth}{@{}>{\raggedright\arraybackslash}p{.205\textwidth}*{8}{Y}@{}}
\toprule
\textbf{Model} & \multicolumn{2}{c}{\textbf{Easy} ($n=138$)} & \multicolumn{2}{c}{\textbf{Medium} ($n=231$)} & \multicolumn{2}{c}{\textbf{Hard} ($n=231$)} & \multicolumn{2}{c}{\textbf{Overall} ($n=600$)} \\
\cmidrule(lr){2-3}\cmidrule(lr){4-5}\cmidrule(lr){6-7}\cmidrule(lr){8-9}
 & P@1 & P@4 & P@1 & P@4 & P@1 & P@4 & P@1 & P@4 \\
\midrule
\rowcolor{sweproprietaryrow}
\multicolumn{9}{c}{\textbf{\emph{Proprietary models}}} \\
\midrule
Claude Opus 4.8 & \textbf{80.4} / \textbf{86.2} & \textbf{84.1} / \textbf{90.6} & 32.9 / 53.3 & 43.7 / 67.0 & \textbf{17.3} / 28.4 & \textbf{22.9} / 39.7 & \textbf{37.8} / 51.3 & \textbf{45.0} / 61.9 \\
\rowcolor{swezebra}
Gemini 3.1 Pro & 56.5 / 70.4 & 64.5 / 79.5 & 14.3 / 46.5 & 23.4 / 60.5 & 2.2 / 25.7 & 8.2 / 37.4 & 19.3 / 44.0 & 27.0 / 55.9 \\
GPT-5.3 Codex & 60.9 / 78.7 & 71.7 / 86.2 & 29.0 / 60.2 & 39.8 / 72.4 & 10.0 / 32.1 & 13.9 / 41.2 & 29.0 / 53.7 & 37.2 / 63.6 \\
\rowcolor{swezebra}
GPT-5.4 & 55.1 / 70.2 & 62.3 / 77.9 & 25.1 / 59.0 & 35.1 / 69.6 & 11.7 / 40.2 & 16.0 / \textbf{48.4} & 26.8 / 54.4 & 34.0 / 63.4 \\
GPT-5.5 & 61.6 / 79.8 & 71.0 / 86.0 & \textbf{35.9} / \textbf{66.0} & \textbf{45.5} / \textbf{77.8} & 14.7 / \textbf{41.5} & 16.9 / 47.1 & 33.7 / \textbf{59.7} & 40.3 / \textbf{67.8} \\
\rowcolor{swezebra}
Qwen3.5-Plus & 42.0 / 54.6 & 60.9 / 73.0 & 6.9 / 42.6 & 21.2 / 59.2 & 1.3 / 21.8 & 2.2 / 29.4 & 12.8 / 37.4 & 23.0 / 50.9 \\
\midrule
\rowcolor{sweopenrow}
\multicolumn{9}{c}{\textbf{\emph{Open-weight models}}} \\
\midrule
DeepSeek-V3.2 & 37.7 / 50.8 & 41.3 / 57.1 & 5.2 / 28.6 & 8.2 / 39.5 & 0.0 / 14.2 & 0.0 / 19.9 & 10.7 / 28.2 & 12.7 / 36.0 \\
\rowcolor{swezebra}
DeepSeek-V4-Flash & 53.6 / 67.8 & 61.6 / 77.8 & 17.3 / 50.7 & 27.3 / 63.4 & 5.2 / 26.9 & 9.5 / 38.6 & 21.0 / 45.4 & 28.3 / 57.2 \\
GLM-5 & 46.4 / 59.4 & 50.7 / 65.4 & 12.1 / 38.8 & 17.7 / 47.6 & 3.9 / 20.2 & 4.3 / 26.1 & 16.8 / 36.3 & 20.2 / 43.4 \\
\rowcolor{swezebra}
GLM-5.2 & 66.7 / 81.4 & 76.1 / 87.1 & 34.6 / 62.9 & \textbf{45.5} / 75.1 & 14.3 / 37.4 & 15.6 / 44.5 & 34.2 / 57.3 & 41.0 / 66.1 \\
Kimi K2.6 & 57.3 / 74.1 & 64.5 / 81.2 & 21.2 / 52.7 & 29.9 / 63.9 & 6.1 / 28.6 & 10.4 / 37.5 & 23.7 / 48.3 & 30.3 / 57.7 \\
\rowcolor{swezebra}
MiniMax-M2.7 & 44.2 / 58.4 & 49.3 / 64.4 & 10.4 / 38.6 & 15.1 / 47.3 & 2.6 / 19.9 & 3.0 / 26.2 & 15.2 / 35.9 & 18.3 / 43.1 \\
\bottomrule
\end{tabularx}
\caption{Claude Code performance by task scope. Cells report resolved rate / scope-weighted score (\%). P@1 and P@4 use one and four attempts. Bold marks metric maxima, including ties.}
\label{tab:difficulty-results}
\end{table*}

\section{Evaluation}
\label{sec:experiments}

\subsection{Evaluation Setup}
\label{sec:eval-setup}

We evaluate 12 proprietary and open-weight model variants on the frozen 600-task benchmark using Claude Code \citep{anthropic2026claudecode}, Cline \citep{cline2026}, and OpenCode \citep{opencode2026}. All models have the same budget of up to 200 turns per scaffold invocation and 6,600 seconds per solver attempt. Each submission is evaluated in a clean environment using the hidden tests and protocol fixed during task construction. Overall and language results use one attempt per task (P@1). The Claude Code analysis by task scope also includes four attempts (P@4). Appendix~\ref{app:run-configurations} lists model references and details execution limits and the handling of timeouts and retries.

We report the \emph{resolved rate}, the percentage of tasks satisfying the strict criterion in Section~\ref{sec:metrics}, and the \emph{scope-weighted score} averaged across tasks and expressed as a percentage. Scoring weights are fixed before solving, following Equation~\ref{eq:weighted} and the configurations in Appendix~\ref{app:metrics}. At P@4, a task is resolved if any of its four patches resolves it, and its scope-weighted score is the maximum achieved by a single patch.

\begin{figure*}[!t]
  \centering
  \includegraphics[width=\textwidth]{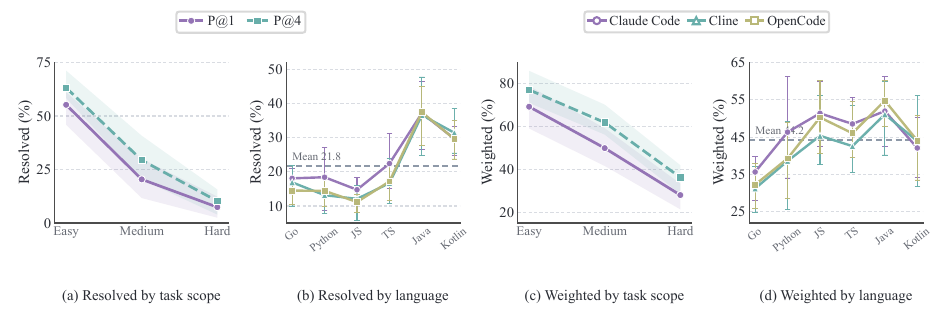}
  \caption{Performance by task scope and language. Panels (a,c) show P@1/P@4 means across 12 models under Claude Code. Panels (b,d) show P@1 means across models for each scaffold. Shading and ranges show the middle 50\%. Dashed lines mark grand means. JS and TS denote JavaScript and TypeScript.}
  \label{fig:performance-overview}
\end{figure*}

\subsection{Main Results}
\label{sec:performance}

Fig.~\ref{fig:overall-performance} compares P@1 performance across all 36 combinations of models and scaffolds. Under Claude Code, Claude Opus 4.8 leads in resolved rate at 37.8\% (51.3\% weighted), whereas GPT-5.5 leads in scope-weighted score at 59.7\% (33.7\% resolved). The leading model therefore differs for complete and partial recovery.

The model with the highest resolved rate also varies by scaffold. Claude Opus 4.8 leads under Claude Code (37.8\%) and OpenCode (29.7\%), while GPT-5.5 and GPT-5.4 tie under Cline (31.2\%).

\subsection{Performance by Task Scope}
\label{sec:scope-results}

Under Claude Code, both P@1 measures decrease from Easy to Medium to Hard for every model (Table~\ref{tab:difficulty-results} and Fig.~\ref{fig:performance-overview}(a,c)). Across the 12 models, the mean resolved rate falls from 55.2\% to 20.4\% and 7.4\%, while the mean scope-weighted score falls from 69.3\% to 50.0\% and 28.1\%.

This ordering persists for every model at P@4. On Hard tasks, four attempts raise the mean resolved rate from 7.4\% to 10.2\% and the scope-weighted score from 28.1\% to 36.3\%. Most Hard tasks remain unresolved under this larger attempt budget. The consistent decline across models and attempt budgets supports the distinction between the three difficulty tiers.

\subsection{Performance across Languages}
\label{sec:language-results}

Under Claude Code (Table~\ref{tab:language-claude-code}), Claude Opus 4.8 attains the highest resolved rate on Go, Python, JavaScript, and TypeScript, while GPT-5.5 leads on Java and Kotlin. For scope-weighted score, GPT-5.5 leads four languages, GPT-5.3 Codex leads TypeScript, and Claude Opus 4.8 leads Java.

% Generated by scripts/plot_results.py; edit the source data or generator.
\begin{table*}[t]
\centering
\fontsize{7.65}{8.65}\selectfont
\renewcommand{\arraystretch}{1.06}
\setlength{\tabcolsep}{1.1pt}
\begin{tabularx}{\textwidth}{@{}>{\raggedright\arraybackslash}p{.225\textwidth}*{6}{Y}@{}}
\toprule
\textbf{Model} & \textbf{Go} & \textbf{Python} & \textbf{JavaScript} & \textbf{TypeScript} & \textbf{Java} & \textbf{Kotlin} \\
\midrule
\rowcolor{sweproprietaryrow}
\multicolumn{7}{c}{\textbf{\emph{Proprietary models}}} \\
\midrule
Claude Opus 4.8 & \textbf{34.0} / 47.1 & \textbf{40.0} / 52.0 & \textbf{30.0} / 53.7 & \textbf{41.0} / 52.5 & 51.0 / \textbf{64.5} & 31.0 / 38.0 \\
\rowcolor{swezebra}
Gemini 3.1 Pro & 16.0 / 37.3 & 8.0 / 37.1 & 12.0 / 49.2 & 16.0 / 43.9 & 35.0 / 51.1 & 29.0 / 45.2 \\
GPT-5.3 Codex & 19.0 / 36.7 & 31.0 / 61.1 & 16.0 / 59.3 & 31.0 / \textbf{57.7} & 45.0 / 60.3 & 32.0 / 46.9 \\
\rowcolor{swezebra}
GPT-5.4 & 14.0 / 36.7 & 27.0 / 61.6 & 19.0 / 63.1 & 26.0 / 57.6 & 44.0 / 57.8 & 31.0 / 49.5 \\
GPT-5.5 & 27.0 / \textbf{49.9} & 24.0 / \textbf{64.9} & 18.0 / \textbf{64.0} & 32.0 / 56.8 & \textbf{54.0} / 64.1 & \textbf{47.0} / \textbf{58.7} \\
\rowcolor{swezebra}
Qwen3.5-Plus & 10.0 / 28.7 & 6.0 / 33.7 & 7.0 / 42.6 & 18.0 / 48.3 & 21.0 / 42.9 & 15.0 / 27.8 \\
\midrule
\rowcolor{sweopenrow}
\multicolumn{7}{c}{\textbf{\emph{Open-weight models}}} \\
\midrule
DeepSeek-V3.2 & 9.0 / 18.5 & 7.0 / 26.8 & 7.0 / 36.8 & 10.0 / 33.8 & 19.0 / 34.5 & 12.0 / 18.7 \\
\rowcolor{swezebra}
DeepSeek-V4-Flash & 18.0 / 34.9 & 11.0 / 41.9 & 11.0 / 48.2 & 16.0 / 46.9 & 33.0 / 47.7 & 37.0 / 53.0 \\
GLM-5 & 14.0 / 26.4 & 11.0 / 34.2 & 12.0 / 42.6 & 12.0 / 39.8 & 27.0 / 41.0 & 25.0 / 34.1 \\
\rowcolor{swezebra}
GLM-5.2 & 27.0 / 49.0 & 27.0 / 61.3 & 22.0 / 61.3 & 34.0 / 55.3 & 52.0 / 64.0 & 43.0 / 53.1 \\
Kimi K2.6 & 17.0 / 37.0 & 20.0 / 47.9 & 13.0 / 53.8 & 22.0 / 50.1 & 40.0 / 55.3 & 30.0 / 46.0 \\
\rowcolor{swezebra}
MiniMax-M2.7 & 12.0 / 25.9 & 9.0 / 33.6 & 10.0 / 41.9 & 11.0 / 39.7 & 25.0 / 40.4 & 24.0 / 34.1 \\
\bottomrule
\end{tabularx}
\caption{Claude Code P@1 performance by language (100 tasks each). Cells report resolved rate / scope-weighted score (\%). Bold marks maxima for each language, including ties.}
\label{tab:language-claude-code}
\end{table*}

Across all 36 combinations, mean P@1 resolved rates range from 12.7\% on JavaScript to 37.1\% on Java, and mean scope-weighted scores range from 33.1\% on Go to 52.6\% on Java (Fig.~\ref{fig:performance-overview}(b,d)). Averaging across models, Claude Code leads four languages on both measures. OpenCode leads on Java for both metrics and on Kotlin for scope-weighted score, while Cline leads on Kotlin for resolved rate. Appendix~\ref{app:evaluation-details} reports results for every model under Cline and OpenCode.

\section{Analysis}
\label{sec:analysis}

\subsection{Interpreting Partial Recovery}

A FlowFuse patch passes 128 of 141 scoring points and achieves a scope-weighted score of 91.6\%. The 13 remaining failures concern getters for stored context, assistant defaults or getters controlled by feature flags, and the format of expert messages (Appendix~\ref{app:cases}). A patch to Jackson's Kotlin module passes 18 of 20 points and scores 90.4\%. Both remain unresolved, but their scope-weighted scores capture substantial partial recovery. For reinforcement learning, this score could provide a graded reward for partial recovery based on the spans of associated function bodies, even when the task remains unresolved. Appendix~\ref{app:metrics} gives the scoring details.

\subsection{Solver Activity in Recorded Trajectories}
\label{sec:behavior}

The corpus contains 7,945 nonempty trajectories from eight models, three agent scaffolds, and six languages. Of these, 7,787 have scorable outcomes. We compare medians for resolved and unresolved attempts within 126 configurations defined by model, scaffold, and language, each with at least five attempts of each outcome.

\begin{figure}[!t]
  \centering
  \includegraphics[width=\columnwidth]{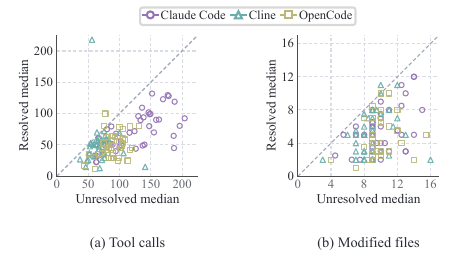}
  \caption{Recorded activity in resolved and unresolved trajectories. Each point compares group medians for one of 126 configurations defined by model, scaffold, and language: (a) tool calls and (b) modified production files. Dashed diagonals mark equality.}
  \label{fig:behavior-comparison}
\end{figure}

% Queue the case figure alongside its analysis on the next page.
% Verified source functions and complete associations: data/composite_task_case.json.
% Rebuild the vector asset with scripts/build_composite_task_graph.py.
\begin{figure*}[!t]
\centering
\includegraphics[width=\textwidth]{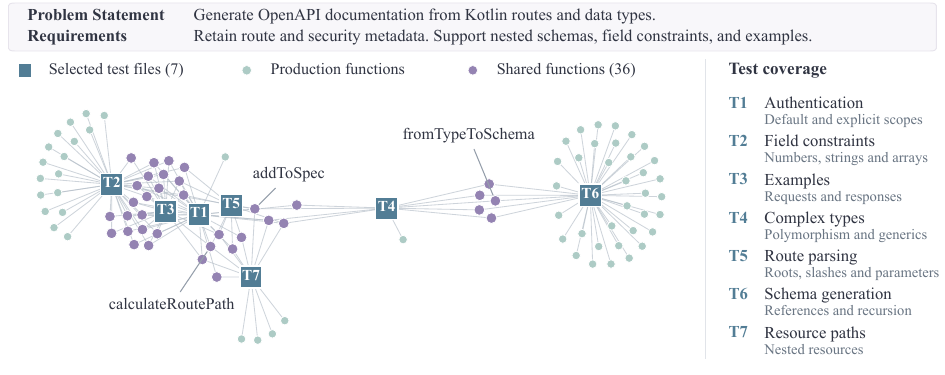}
\caption{A composite task for OpenAPI documentation generation in Kompendium, with an abridged description. Seven selected test files cover 94 production functions across 33 source files. Purple circles mark the 36 functions associated with multiple selected test files.}
\label{fig:composite-task-case}
\end{figure*}

Fig.~\ref{fig:behavior-comparison} shows lower medians for resolved attempts in 122 of 126 configurations for tool calls (96.8\%) and in 123 for modified files (97.6\%). The median differences (resolved minus unresolved) are $-35.5$ calls and $-4$ files, with the same pattern in most configurations under each scaffold. Task difficulty and stopping after completion may partly explain this association.

\subsection{Case Study: Task Composition}
\label{sec:composition-case}

Fig.~\ref{fig:composite-task-case} illustrates a composite task for generating OpenAPI documentation from Kotlin routes and data types. Seven test files combine requirements for authentication, field constraints, examples, complex types, route parsing, schema generation, and nested resources. The tests check generated schemas and the served OpenAPI document against expected JSON snapshots. Composition expands the implementation scope from at most 40 functions in any individual footprint to 94 functions. Shared functions such as \texttt{addToSpec} and \texttt{fromTypeToSchema} connect document assembly with routing and type conversion. All 39 expected scoring points pass originally and fail under the complete mask, affecting every selected file.

% Allow text to continue while the double-column case figure is queued.
\section{Conclusion}
\label{sec:conclusion}

\method{} is a fully automated, four-agent pipeline for constructing repository-level tasks from test execution. It validates candidate scopes against original and masked repositories, then rechecks every constituent test file after deterministic composition. Evaluation applies the same hidden tests and fixed protocol across submissions. The benchmark contains 600 tasks from 105 repositories across six languages. Across 12 models and three agent scaffolds, the best configuration resolves 37.8\% of tasks in one attempt. Scope-weighted scoring captures partial recovery with weights fixed from observed test--function associations before solving. Composing validated execution footprints extends tasks beyond individual historical changes to span functions and modules.

\section*{Limitations}

\method{} relies on existing repository tests to identify implementation scopes and validate task behavior. Behaviors and edge cases that these tests do not exercise may be underrepresented in the resulting tasks. Extending the approach to functionality with limited test coverage requires developing and validating additional tests. The benchmark currently covers six programming languages. Broader language coverage requires adapting tracing, source mapping, and result parsing to additional runtimes and test frameworks.

% Avoid leaving a single final line of the ethics paragraph on the next page.
\begingroup
\widowpenalty=10000
\brokenpenalty=10000
\section*{Ethical Considerations}

This benchmark derives implementation tasks from public software repositories. Use and redistribution of repository code, tests, and derived tasks remain subject to the original licenses. During construction and evaluation, code from repositories and agents should run in environments isolated from sensitive hosts and credentials, since tests and setup commands can have side effects. Benchmark results verify selected test outcomes and do not establish security or deployment readiness. The study does not involve human participants.

\par
\endgroup
\bibliography{references}
\appendix
\sweappendixfloats
\raggedbottom
% Let the tracing table float while the construction details continue in two columns.
\section{Execution and Task Construction Details}
\label{app:execution}

\begin{table*}[!bp]
\centering
\small
\begin{tabular}{p{1.65cm}p{6.45cm}p{6.45cm}}
\toprule
Language & Representative observation mechanism & Scope or mapping constraint \\
\midrule
Python & Profiling callbacks during pytest execution, with source paths and qualified names. & A call-phase hook does not automatically cover setup, teardown, native extensions, or subprocesses. \\
Go & Function-entry instrumentation in source files supplied through a build overlay. & Package compilation and execution can exceed the selected physical test file. \\
JavaScript & Source instrumentation or test-runner transforms in the executing Node.js process. & Export-only instrumentation and worker coverage depend on the chosen backend. \\
TypeScript & Source-aware transforms integrated with Vite/Vitest or the repository runner. & Transpilation, expression bodies, and test ownership require explicit source mapping. \\
Java & Source probes or runtime bytecode instrumentation with test-framework hooks. & Constructors, generated methods, and thread boundaries affect observation and localization. \\
Kotlin & Source probes or runtime observation followed by mapping to Kotlin declarations. & Synthetic wrappers and declarations without independent source bodies require special handling. \\
\bottomrule
\end{tabular}
\caption{Representative tracing mechanisms and coverage constraints. The mechanism is adapted to each repository's test runner and source layout.}
\label{tab:tracing}
\end{table*}

\paragraph{Repository adaptation.}
Tracing examples provide instrumentation strategies and repository instructions. The Trace Agent adapts these materials to each selected test, defining its execution command and process scope. Table~\ref{tab:tracing} summarizes the supported mechanisms and their coverage constraints.

\paragraph{Structural masking.}
The masking procedure locates editable bodies in clean source independently of the locations reported by tracing. We use source identities to deduplicate aliases and shared functions. Language-specific stubs preserve declarations while making missing behavior observable. Adjustments include retaining explicit constructor delegation in Java, adapting unused imports in Go, and preserving necessary type relationships in TypeScript. Structural checks compare the original and masked source and verify that the masking patch applies correctly. Independent runs on both repositories check the acceptance conditions in Equation~\ref{eq:admission}.

\paragraph{Execution protocol.}
The evaluation protocol specifies environment setup, test execution, report collection, and result parsing for both repository states in clean environments. It also records exit status and test identifiers. The Validation Agent may refine the parser before acceptance, and any subsequent change requires renewed validation. When outcomes are available only for a test file as a whole, validation and evaluation use the same file-level scoring unit. Missing test case reports are not counted as individual failures.

\paragraph{Composition.}
For two observed execution footprints $A$ and $B$, overlap is measured by $J(A,B)=\ell(A\cap B)/\ell(A\cup B)$, where $\ell(U)=\sum_{f\in U}L_f$ and $L_f$ is the recorded body span of function $f$ in source lines. This ratio weights functions by their body spans. Candidate groups can be selected by union size and overlap. Every constituent test file must pass validation under the complete combined mask. Each scoring point is identified by its source test file and original test identifier to avoid collisions.

\paragraph{Specification and access policies.}
The Query Agent uses the original code, selected tests, repository context, and specification examples to generate a Problem Statement and Requirements. Function declarations and signatures remain available in the masked repository. Checks for format and literal leakage run before packaging. We remove the original Git object history and withhold construction artifacts from the solver workspace. Task instructions restrict external retrieval of repository source or solution code. Appendix~\ref{app:cases} gives the specification format and recorded examples.

% Ordinary prose stays in two columns; the case displays float across the page.
\section{Task Specifications and Case Studies}
\label{app:cases}

Fig.~\ref{fig:prompt-contract} details the specification and solver inputs introduced in Fig.~\ref{fig:method}.

The examples use the Problem Statement and Requirements fields and the same execution restrictions. Some prompts also repeat function signatures available in the masked repository. We present atomic and composite tasks in Fig.~\ref{fig:task-form-cases} and recorded solver actions with independent evaluation outcomes in Fig.~\ref{fig:trajectory-cases}. Appendix~\ref{app:task-examples} provides detailed specifications adapted from six composite tasks.

\begin{figure*}[!t]
\centering
\begin{minipage}[t]{0.485\textwidth}
\begin{swepromptbox}[equal height group=promptcontract]{Specification generation (abridged)}
\textbf{Evidence.} Original production bodies, selected tests, and repository context are provided to the Query Agent.

\smallskip
\textbf{Goal.} Write a self-contained behavioral specification for the missing implementation scope.

\smallskip
\textbf{Coverage.} State observable inputs and outputs, boundary cases, defaults, errors, and side effects exercised by the selected tests.

\smallskip
\textbf{Constraints.} Do not reveal reference code, copy wording from hidden tests, or prescribe an implementation strategy.

\smallskip
\textbf{Output.} Provide a Problem Statement and Requirements describing the behavior to recover.
\end{swepromptbox}
\end{minipage}
\hfill
\begin{minipage}[t]{0.485\textwidth}
\begin{swepromptbox}[equal height group=promptcontract]{Solver input}
\texttt{PROBLEM STATEMENT: <behavioral overview>}

\smallskip
\texttt{REQUIREMENTS: 1. <observable behavior>; 2. <boundary case>; ...}

\smallskip
\texttt{<benchmark integrity policy>}

\medskip
\textbf{Workspace.} The masked repository accompanies this text. The evaluator retains the selected tests, construction traces, reference bodies, and scoring metadata.
\end{swepromptbox}
\end{minipage}
\caption{Specification generation and solver inputs. The Query Agent uses the original implementation and selected tests to describe behavior. The solver receives the specification and masked repository, which retains function declarations and signatures.}
\label{fig:prompt-contract}
\end{figure*}

\subsection{Concrete Task Forms}

Fig.~\ref{fig:task-form-cases} gives one released instance for each task form. An atomic task uses one selected test file and may mask multiple functions. A composite task combines the validated scopes of multiple test files, counting shared functions once.

\begin{figure*}[!t]
\centering
\begin{minipage}[t]{0.485\textwidth}
\begin{swecasebox}[equal height group=taskforms,colbacktitle=sweteal]{Atomic task: eclipse/hono (Java)}
\textbf{Selected behavior.} The point in \texttt{HttpUtilsTest.java} checks that an invalid Uniform Resource Identifier (URI) still yields a non-null absolute URI.

\smallskip
\textbf{Task scope.} One selected test file, one failure-to-pass scoring point, and 15 changed lines in the reference restoration patch.

\smallskip
\textbf{Task boundary.} Validation checks that the selected test passes in the original repository and fails or reports an error when the implementation scope is masked.
\end{swecasebox}
\end{minipage}
\hfill
\begin{minipage}[t]{0.485\textwidth}
\begin{swecasebox}[equal height group=taskforms]{Composite task: canonical/cloud-init (Python)}
\textbf{Selected behavior.} The task combines hosts file operations and resolver configuration from \texttt{test\_hosts.py} and \texttt{test\_resolv.py}.

\smallskip
\textbf{Task scope.} The combined scope contains 22 masked bodies spanning 216 lines and seven expected scoring points. The reference restoration patch changes 208 lines.

\smallskip
\textbf{Task boundary.} The two parent tests share observed functions and are both revalidated under the complete combined mask.
\end{swecasebox}
\end{minipage}
\caption{Atomic and composite tasks with their validated implementation scopes.}
\label{fig:task-form-cases}
\end{figure*}

\subsection{Recorded Execution Cases}

Fig.~\ref{fig:trajectory-cases} contrasts a fully resolved task with a high-scoring patch that leaves several behaviors incomplete. The excerpts include only visible tool actions and recorded results.

\begin{figure*}[!t]
\centering
\begin{swecasebox}[colbacktitle=sweteal]{Case 1: canonical/cloud-init (resolved composite)}
\begin{minipage}[t]{0.275\textwidth}
\textbf{Task.} Recover host and resolver parsing, queries, mutation, rendering, stable deduplication, and resolver limits.

\smallskip
\textbf{Scope.} 2 selected test files, 22 masked bodies, and 216 body lines.

\smallskip
\textbf{Recorded work.} 47 tool calls and 4 modified production files.
\end{minipage}
\hfill
\begin{minipage}[t]{0.675\textwidth}
\textbf{Observable trajectory.}
\begin{enumerate}[leftmargin=1.45em,itemsep=2pt,topsep=3pt,parsep=0pt,label=\textcolor{sweteal}{\bfseries\arabic*.}]
  \item Search and read the parser modules, then implement the host, resolver, and shared helper behavior.
  \item Run focused checks successfully. A broader distribution suite then reports 14 failures and 12 errors.
  \item Inspect the remaining failures and locate a second masked \texttt{uniq\_list} dependency in \texttt{cloudinit/util.py}.
  \item Repair that helper and rerun the broader checks. The distribution and network suites each report 384 passing tests.
\end{enumerate}
\smallskip\hrule\smallskip
\textbf{Independent evaluation.} Restoring the two hidden test files yields 7/7 failure-to-pass points, complete resolution, and a scope-weighted score of 1.000.
\end{minipage}
\end{swecasebox}

\vspace{7pt}

\begin{swecasebox}{Case 2: FlowFuse/flowfuse (partial recovery)}
\begin{minipage}[t]{0.275\textwidth}
\textbf{Task.} Recover device-group requests, authentication and context state, assistant feature gates, expert-assistant state, and timer utilities.

\smallskip
\textbf{Scope.} 6 selected test files, 85 masked bodies, and 646 body lines.

\smallskip
\textbf{Recorded work.} 71 tool calls and 8 modified production files.
\end{minipage}
\hfill
\begin{minipage}[t]{0.675\textwidth}
\textbf{Observable trajectory.}
\begin{enumerate}[leftmargin=1.45em,itemsep=2pt,topsep=3pt,parsep=0pt,label=\textcolor{sweviolet}{\bfseries\arabic*.}]
  \item Read the application client, timer, and store modules, then organize the specification into eight implementation units.
  \item Edit the corresponding production modules and install the repository dependencies needed for verification.
  \item Run the visible Vitest configuration, which reports 29 test files and 351 tests passing.
\end{enumerate}
\smallskip\hrule\smallskip
\textbf{Independent evaluation.} After the six hidden test files are restored, the patch passes 128/141 points. The 13 omissions comprise five context-store getters, seven assistant defaults or feature-gated getters, and one expert-message format. The scope-weighted score is 0.916, but the task remains unresolved.
\end{minipage}
\end{swecasebox}
\caption{Recorded tool actions and outcomes from two trajectories. Visible checks guide repair, and independent evaluation determines complete and partial recovery.}
\label{fig:trajectory-cases}
\end{figure*}

\subsection{Cases across Languages}

Table~\ref{tab:case-portfolio} reports first-attempt trajectories across five languages with the model and scaffold fixed. The cases cover all three scope tiers and include fully resolved tasks, partial recovery, and zero recovery.

\begin{table*}[!t]
\centering
\fontsize{8.1}{9.2}\selectfont
\setlength{\tabcolsep}{3.2pt}
\renewcommand{\arraystretch}{1.05}
\begin{tabularx}{\textwidth}{@{}l>{\raggedright\arraybackslash}Xlrrrrr@{}}
\toprule
\rowcolor{sweopenrow}
Language & Repository & Tier & $k$ & Functions / lines & Calls & Resolved & Weighted \\
\midrule
\rowcolor{swezebra}
Python & canonical/cloud-init & Easy & 2 & 22 / 216 & 47 & Yes & 100.0 \\
JavaScript & FlowFuse/flowfuse & Medium & 6 & 85 / 646 & 71 & No & 91.6 \\
\rowcolor{swezebra}
TypeScript & minvws/NL COVID-19 Dashboard & Easy & 2 & 4 / 113 & 50 & Yes & 100.0 \\
Java & apache/avro & Hard & 7 & 632 / 5,198 & 89 & No & 0.0 \\
\rowcolor{swezebra}
Kotlin & FasterXML/Jackson Kotlin & Medium & 6 & 49 / 553 & 192 & No & 90.4 \\
\bottomrule
\end{tabularx}
\caption{First-attempt Qwen3.5-Plus cases under Claude Code across five languages. $k$ counts selected test files. Functions / lines gives unique functions and recorded body spans. Weighted is the scope-weighted score (\%).}
\label{tab:case-portfolio}
\end{table*}

% Pair the compact statistics tables in one full-width float.
\section{Dataset Scope and Statistics}
\label{app:data}

\begin{table*}[!t]
\begin{sweappendixblock}{0.485\textwidth}
% Generated by scripts/plot_results.py; edit the source data or generator.
\begin{table}[t]
\centering
\small
\setlength{\tabcolsep}{3pt}
\begin{tabular}{lrrrr}
\toprule
Language & Tasks & Repos. & Mean $N$ & Median $N$ \\
\midrule
Go & 100 & 28 & 14.68 & 9 \\
Python & 100 & 18 & 35.68 & 26 \\
JavaScript & 100 & 13 & 35.66 & 21 \\
TypeScript & 100 & 19 & 31.67 & 25 \\
Java & 100 & 13 & 27.15 & 14 \\
Kotlin & 100 & 14 & 14.91 & 11.5 \\
\midrule
All & 600 & 105 & 26.63 & 17 \\
\bottomrule
\end{tabular}
\caption{Dataset statistics by language. $N$ counts scoring points per task. Repository counts are per language. Repositories and scoring points may recur across tasks.}
\label{tab:dataset}
\end{table}

\end{sweappendixblock}\hfill
\begin{sweappendixblock}{0.485\textwidth}
\begin{table}[t]
\centering
\small
\setlength{\tabcolsep}{2.2pt}
\begin{tabular}{lrrrr}
\toprule
Statistic & Easy & Medium & Hard & All \\
\midrule
Tasks & 138 & 231 & 231 & 600 \\
Selected test files (median) & 2 & 4 & 8 & 5 \\
Masked bodies (median) & 11 & 30 & 119 & 35 \\
Reference-patch files (median) & 3 & 9 & 20 & 10 \\
Changed lines (median) & 109.5 & 418 & 1,555 & 433 \\
Scoring points (median) & 5 & 16 & 28 & 17 \\
\bottomrule
\end{tabular}
\caption{Recorded task scope by tier and overall. Masked bodies are production function bodies. File and line counts describe reference restoration patches. Changed lines count additions and deletions and differ from the source spans used for weighting.}
\label{tab:tier-statistics}
\end{table}
\end{sweappendixblock}
\end{table*}

\paragraph{Dataset composition.}
Tables~\ref{tab:dataset} and~\ref{tab:tier-statistics} summarize the frozen 600-task release by language and scope tier. The numbers of tasks selecting one through nine test files are respectively 116, 56, 59, 61, 61, 67, 63, 55, and 62. We link test files that share observed production functions. Of the 484 composite tasks, 255 have connected graphs with one component, and 229 have disconnected graphs with multiple components. Every constituent test file is validated under the complete combined mask, regardless of graph structure.

\paragraph{Repository and parent reuse.}
The largest repository contributes 24\% of Go tasks, 36\% of Python, 23\% of JavaScript, 18\% of TypeScript, 27\% of Java, and 28\% of Kotlin. A validated parent test may participate in multiple composites. The maximum reuse in these language sets is 8, 24, 16, 12, 21, and 19 tasks, respectively. Task outcomes can therefore be statistically dependent, and future training and evaluation splits should control shared repositories and parent tests.

\section{Evaluation Details}
\label{app:evaluation-details}

\subsection{Evaluation Configuration}
\label{app:run-configurations}

Table~\ref{tab:model-references} lists the evaluated model variants and their reference sources. All configurations use the common execution limits in Table~\ref{tab:run-configurations}. The output limit applies to each model request, and the solver deadline applies to each attempt. The Claude Code and OpenCode versions are 2.1.158 and 1.3.17, respectively.

% Official release references; model labels match the frozen result tables.
\begin{table}[!htbp]
\centering
\small
\setlength{\tabcolsep}{3pt}
\begin{tabularx}{\columnwidth}{@{}l>{\raggedright\arraybackslash}X@{}}
\toprule
Model & Reference \\
\midrule
Claude Opus 4.8 & \citep{anthropic2026opus48} \\
\rowcolor{swezebra}
Gemini 3.1 Pro & \citep{google2026gemini31pro} \\
GPT-5.3 Codex & \citep{openai2026gpt53codex} \\
\rowcolor{swezebra}
GPT-5.4 & \citep{openai2026gpt54} \\
GPT-5.5 & \citep{openai2026gpt55} \\
\rowcolor{swezebra}
Qwen3.5-Plus & \citep{qwen2026qwen35} \\
DeepSeek-V3.2 & \citep{deepseek2025v32} \\
\rowcolor{swezebra}
DeepSeek-V4-Flash & \citep{deepseek2026v4flash} \\
GLM-5 & \citep{glm2026glm5} \\
\rowcolor{swezebra}
GLM-5.2 & \citep{zai2026glm52} \\
Kimi K2.6 & \citep{moonshot2026kimik26} \\
\rowcolor{swezebra}
MiniMax-M2.7 & \citep{minimax2026m27} \\
\bottomrule
\end{tabularx}
\caption{Reference sources for the evaluated model variants.}
\label{tab:model-references}
\end{table}

\begin{table}[!htbp]
\centering
\small
\setlength{\tabcolsep}{4pt}
\begin{tabularx}{\columnwidth}{@{}>{\raggedright\arraybackslash}X>{\raggedright\arraybackslash}X@{}}
\toprule
Setting & Value \\
\midrule
Agent turns & 200 per scaffold invocation \\
\rowcolor{swezebra}
Solver time limit & 6,600 seconds per attempt \\
Runtime timeout & 7,200 seconds \\
\rowcolor{swezebra}
Output token limit & 32,000 per model request \\
\bottomrule
\end{tabularx}
\caption{Common execution limits across all evaluated models and agent scaffolds. These limits are fixed before solving.}
\label{tab:run-configurations}
\end{table}

On timeout, the runner preserves the current production patch for independent evaluation. Infrastructure retries are recorded separately from the solver attempts used for P@1/P@4. Evaluation applies the same hidden tests, parsing rules, and scoring criteria to every submission for a given task.

\subsection{Complete Results}

% Generated by scripts/plot_results.py; edit the source data or generator.
\begin{table}[!t]
\centering
\fontsize{6.75}{7.75}\selectfont
\renewcommand{\arraystretch}{1.06}
\setlength{\tabcolsep}{0.7pt}
\begin{tabularx}{\columnwidth}{@{}>{\raggedright\arraybackslash}p{.37\columnwidth}*{3}{Y}@{}}
\toprule
\textbf{Model} & \textbf{Claude Code} & \textbf{Cline} & \textbf{OpenCode} \\
\midrule
\rowcolor{sweproprietaryrow}
\multicolumn{4}{c}{\textbf{\emph{Proprietary models}}} \\
\midrule
Claude Opus 4.8 & \textbf{37.8} / 51.3 & 26.5 / 43.5 & \textbf{29.7} / \textbf{53.9} \\
\rowcolor{swezebra}
Gemini 3.1 Pro & 19.3 / 44.0 & 19.5 / 43.0 & 18.2 / 41.4 \\
GPT-5.3 Codex & 29.0 / 53.7 & 27.7 / 52.4 & 27.8 / 53.8 \\
\rowcolor{swezebra}
GPT-5.4 & 26.8 / 54.4 & \textbf{31.2} / 55.4 & 23.8 / 51.3 \\
GPT-5.5 & 33.7 / \textbf{59.7} & \textbf{31.2} / \textbf{58.0} & 25.5 / 53.6 \\
\rowcolor{swezebra}
Qwen3.5-Plus & 12.8 / 37.4 & 9.3 / 22.1 & 13.2 / 33.2 \\
\midrule
\rowcolor{sweopenrow}
\multicolumn{4}{c}{\textbf{\emph{Open-weight models}}} \\
\midrule
DeepSeek-V3.2 & 10.7 / 28.2 & 13.2 / 30.2 & 13.5 / 31.9 \\
\rowcolor{swezebra}
DeepSeek-V4-Flash & 21.0 / 45.4 & 13.5 / 34.5 & 16.5 / 41.0 \\
GLM-5 & 16.8 / 36.3 & 14.8 / 32.8 & 16.5 / 36.6 \\
\rowcolor{swezebra}
GLM-5.2 & 34.2 / 57.3 & 30.3 / 54.4 & 26.5 / 53.7 \\
Kimi K2.6 & 23.7 / 48.3 & 23.2 / 47.2 & 22.5 / 47.0 \\
\rowcolor{swezebra}
MiniMax-M2.7 & 15.2 / 35.9 & 13.3 / 32.1 & 14.8 / 36.0 \\
\bottomrule
\end{tabularx}
\caption{Exact overall P@1 performance on \method{} (\%). Cells report resolved rate / scope-weighted score. Bold marks metric maxima, including ties.}
\label{tab:main-results}
\end{table}

Tables~\ref{tab:main-results} and~\ref{tab:language-other-scaffolds} report complete P@1 results, rounded to one decimal. Means are computed from tabulated values over the models or configurations compared.

% Generated by scripts/plot_results.py; edit the source data or generator.
\begin{table*}[!t]
\centering
\fontsize{7.65}{8.65}\selectfont
\renewcommand{\arraystretch}{1.06}
\setlength{\tabcolsep}{1.1pt}
\begin{tabularx}{\textwidth}{@{}>{\raggedright\arraybackslash}p{.225\textwidth}*{6}{Y}@{}}
\toprule
\textbf{Model} & \textbf{Go} & \textbf{Python} & \textbf{JavaScript} & \textbf{TypeScript} & \textbf{Java} & \textbf{Kotlin} \\
\midrule
\rowcolor{sweproprietaryrow}
\multicolumn{7}{c}{\textbf{Cline}\enspace\textperiodcentered\enspace\textbf{\emph{Proprietary models}}} \\
\midrule
Claude Opus 4.8 & \textbf{28.0} / 39.4 & 20.0 / 47.2 & 21.0 / 38.9 & 22.0 / 41.2 & 40.0 / 56.2 & 28.0 / 38.2 \\
\rowcolor{swezebra}
Gemini 3.1 Pro & 20.0 / 33.5 & 11.0 / 39.1 & 13.0 / 44.4 & 11.0 / 40.5 & 36.0 / 53.9 & 26.0 / 46.7 \\
GPT-5.3 Codex & 16.0 / 32.8 & \textbf{24.0} / 47.6 & 14.0 / 59.1 & 24.0 / 53.2 & 51.0 / 64.2 & 37.0 / 57.6 \\
\rowcolor{swezebra}
GPT-5.4 & 20.0 / 36.4 & 21.0 / \textbf{56.3} & \textbf{25.0} / \textbf{64.8} & 24.0 / 53.9 & 47.0 / 58.3 & \textbf{50.0} / \textbf{62.4} \\
GPT-5.5 & 27.0 / \textbf{43.2} & 16.0 / 55.7 & 16.0 / 60.5 & \textbf{29.0} / \textbf{60.8} & \textbf{53.0} / \textbf{66.7} & 46.0 / 61.4 \\
\rowcolor{swezebra}
Qwen3.5-Plus & 7.0 / 14.5 & 4.0 / 18.9 & 5.0 / 20.8 & 0.0 / 21.2 & 23.0 / 31.0 & 17.0 / 26.1 \\
\midrule
\rowcolor{sweopenrow}
\multicolumn{7}{c}{\textbf{Cline}\enspace\textperiodcentered\enspace\textbf{\emph{Open-weight models}}} \\
\midrule
DeepSeek-V3.2 & 9.0 / 19.8 & 9.0 / 25.1 & 6.0 / 36.6 & 15.0 / 39.2 & 23.0 / 36.0 & 17.0 / 24.7 \\
\rowcolor{swezebra}
DeepSeek-V4-Flash & 10.0 / 30.5 & 4.0 / 25.7 & 5.0 / 38.1 & 6.0 / 28.7 & 25.0 / 44.2 & 31.0 / 39.6 \\
GLM-5 & 11.0 / 24.8 & 8.0 / 25.9 & 6.0 / 37.7 & 12.0 / 35.8 & 26.0 / 40.4 & 26.0 / 31.9 \\
\rowcolor{swezebra}
GLM-5.2 & 27.0 / 42.3 & 18.0 / 53.6 & 16.0 / 55.1 & 28.0 / 55.9 & 50.0 / 64.1 & 43.0 / 55.6 \\
Kimi K2.6 & 18.0 / 33.2 & 16.0 / 42.9 & 13.0 / 51.0 & 19.0 / 46.2 & 41.0 / 58.5 & 32.0 / 51.6 \\
\rowcolor{swezebra}
MiniMax-M2.7 & 10.0 / 24.6 & 7.0 / 25.4 & 5.0 / 37.3 & 10.0 / 34.5 & 24.0 / 39.7 & 24.0 / 31.4 \\
\addlinespace[2pt]
\rowcolor{sweproprietaryrow}
\multicolumn{7}{c}{\textbf{OpenCode}\enspace\textperiodcentered\enspace\textbf{\emph{Proprietary models}}} \\
\midrule
Claude Opus 4.8 & \textbf{26.0} / \textbf{43.6} & \textbf{27.0} / \textbf{55.8} & \textbf{18.0} / \textbf{66.9} & \textbf{31.0} / 54.9 & 43.0 / 57.0 & 33.0 / 45.4 \\
\rowcolor{swezebra}
Gemini 3.1 Pro & 16.0 / 33.4 & 9.0 / 34.9 & 9.0 / 41.7 & 7.0 / 33.6 & 39.0 / 56.1 & 29.0 / 48.6 \\
GPT-5.3 Codex & 19.0 / 37.4 & 18.0 / 46.0 & 13.0 / 59.4 & 26.0 / 54.2 & 48.0 / 63.7 & \textbf{43.0} / \textbf{62.2} \\
\rowcolor{swezebra}
GPT-5.4 & 9.0 / 30.2 & 23.0 / 54.6 & 13.0 / 59.4 & 24.0 / \textbf{56.0} & 42.0 / 57.4 & 32.0 / 50.5 \\
GPT-5.5 & 17.0 / 39.8 & 15.0 / 48.6 & 15.0 / 62.3 & 21.0 / 54.5 & \textbf{50.0} / \textbf{64.4} & 35.0 / 51.8 \\
\rowcolor{swezebra}
Qwen3.5-Plus & 3.0 / 23.2 & 8.0 / 27.5 & 7.0 / 37.4 & 16.0 / 34.7 & 24.0 / 43.5 & 21.0 / 32.6 \\
\midrule
\rowcolor{sweopenrow}
\multicolumn{7}{c}{\textbf{OpenCode}\enspace\textperiodcentered\enspace\textbf{\emph{Open-weight models}}} \\
\midrule
DeepSeek-V3.2 & 9.0 / 20.1 & 11.0 / 28.7 & 7.0 / 34.3 & 8.0 / 32.6 & 22.0 / 41.9 & 24.0 / 34.0 \\
\rowcolor{swezebra}
DeepSeek-V4-Flash & 15.0 / 32.0 & 8.0 / 27.6 & 8.0 / 47.6 & 14.0 / 51.7 & 34.0 / 55.3 & 20.0 / 32.1 \\
GLM-5 & 13.0 / 26.1 & 12.0 / 29.2 & 10.0 / 40.8 & 12.0 / 41.6 & 28.0 / 48.0 & 24.0 / 33.7 \\
\rowcolor{swezebra}
GLM-5.2 & 20.0 / 40.8 & 18.0 / 50.4 & 15.0 / 63.5 & 22.0 / 54.6 & 48.0 / 62.5 & 36.0 / 50.2 \\
Kimi K2.6 & 16.0 / 35.2 & 14.0 / 39.9 & 11.0 / 49.7 & 15.0 / 42.9 & 44.0 / 59.5 & 35.0 / 54.7 \\
\rowcolor{swezebra}
MiniMax-M2.7 & 11.0 / 25.4 & 10.0 / 28.2 & 8.0 / 40.2 & 10.0 / 41.2 & 27.0 / 47.9 & 23.0 / 33.1 \\
\bottomrule
\end{tabularx}
\caption{Cline and OpenCode P@1 performance by language (100 tasks each). Cells report resolved rate / scope-weighted score (\%). Bold marks maxima for each language and scaffold, including ties.}
\label{tab:language-other-scaffolds}
\end{table*}

\section{Scoring and Trajectory Analysis}
\subsection{Scope-Weighted Scoring}
\label{app:metrics}

Equation~\ref{eq:weighted} divides each function's recorded span equally among its associated scoring points, giving $\sum_p w_p=\sum_f L_f$. Coarser attribution associates a function with all points in its execution record. A file supplies one point when individual outcomes are unavailable. In the release, 573 tasks assign positive weight to every expected point. All scoring configurations, including coarser associations or unit weights where finer attribution is unavailable, are fixed before solving.

As a synthetic example, suppose a 20-line function is associated with points $p_1$ and $p_2$, a 30-line function only with $p_2$, and a 50-line function only with $p_3$. Their point weights are 10, 40, and 50. Passing $p_1$ and $p_2$ yields an unweighted score of $2/3$ and a scope-weighted score of $0.5$, but the task remains unresolved.

In recorded attempts, a FlowFuse patch passes 128 of 141 points with a scope-weighted score of 0.915988, and a patch to Jackson's Kotlin module passes 18 of 20 with a score of 0.904475. Both remain unresolved. Appendix~\ref{app:cases} traces the FlowFuse attempt from visible verification to independent evaluation.

\subsection{Trajectory Analysis}
\label{app:behavior}

Each trajectory is identified by its model, scaffold, task instance, and trajectory hash. Of 7,945 nonempty trajectories, 7,787 are scorable. We compare median activity between resolved and unresolved attempts for each combination of model, scaffold, and language. Of 144 configurations, 126 contain at least five attempts of each outcome and are retained.

Tool call counts measure visible activity, and modified file counts measure the breadth of code changes. We also record whether the solver verifies its edits and where the first edit occurs in the trajectory. Across configurations, the median differences (resolved minus unresolved) are $+4.39$ and $+2.11$ percentage points, respectively.

% Native two-column analysis precedes six full-width task-specification figures.
% Original prompts, task identifiers, and source hashes are retained in
% data/appendix_task_examples.json.
\section{Task Examples across Six Languages}
\label{app:task-examples}

Figs.~\ref{fig:task-example-go}--\ref{fig:task-example-kotlin} present one composite task for each language in \method. We select tasks that combine complementary constraints on a coherent software feature. The specifications preserve their functional goals, boundary cases, and errors, with wording edited for presentation.

\paragraph{Go: Broker lifecycle.} Fig.~\ref{fig:task-example-go} combines six selected test files into one Broker lifecycle. Contract contents, readiness, and cleanup must remain consistent. Reconciliation must avoid unnecessary generation increments, while finalization preserves external topics.

\paragraph{Python: REST API runtime.} Fig.~\ref{fig:task-example-python} combines seven selected test files spanning request parsing and response representations. Form parsing must preserve the underlying Django request, and version selection must agree with hyperlink generation and validation. Distinguishing parser errors from ordinary attribute access adds another constraint across the framework's request, serializer, and URL components.

\paragraph{JavaScript: Tour lifecycle.} Fig.~\ref{fig:task-example-js} combines nine selected test files covering rendered components, events, steps, and attachment utilities. Together they describe a complete user-facing interaction, including asynchronous showing and later cleanup. A step may render correctly on its first appearance yet mishandle a changed target or overwrite its original tabindex on a later show. The example makes lifecycle consistency and reversible DOM changes part of the same task.

\paragraph{TypeScript: Tab interaction.} Fig.~\ref{fig:task-example-ts} combines seven selected test files around Tabs and its supporting utilities. The task connects asynchronous guards, public event ordering, keyboard focus, and reactive rendering. An immediate click event must coexist with a delayed state update, while reordered children must update both header order and indicator position. The example brings component state, event semantics, and supporting utilities under one behavioral specification.

\paragraph{Java: Schema evolution.} Fig.~\ref{fig:task-example-java} combines eight selected test files around a shared schema model. The feature spans construction, compatibility checking, and runtime use: accepting a default in a builder is insufficient if decoding later produces different bytes. Recursive histories and precise incompatibility locations add further constraints. The example illustrates a composite task in which several APIs must implement the same data-model rules consistently.

\paragraph{Kotlin: Multipart integrity.} Fig.~\ref{fig:task-example-kotlin} combines nine selected test files covering object request conversion and multipart integrity. The example links accepted header formats to upload validation, decoded payloads, and digest generation. A matching ETag does not make an incorrectly ordered completion valid, and a correctly decoded body must still yield the required checksum representation. Together, these requirements specify object request handling and multipart integrity with concrete success and error behavior.

\tcbset{swetaskexample/.style={
  colback=white,colframe=sweviolet!40,
  colbacktitle=sweviolet!16,coltitle=sweviolet!45!black,
  boxrule=0.5pt,arc=2pt,
  left=8pt,right=8pt,top=6pt,bottom=6pt,
  toptitle=5pt,bottomtitle=5pt,
  fonttitle=\bfseries\fontsize{11}{13}\selectfont,
  fontupper=\fontsize{9.5}{11.6}\selectfont\raggedright
}}
\newenvironment{swetaskrequirements}[1]{%
  \begin{enumerate}[start=#1,leftmargin=1.65em,labelsep=0.45em,
    itemsep=5pt,topsep=4pt,parsep=0pt,
    label=\textcolor{sweviolet!75!black}{\bfseries\arabic*.}]
  \raggedright
}{\end{enumerate}}

% The first case can fill the space below the final analysis paragraph.
% Subsequent full-width cases retain the native two-column appendix layout.
\begin{figure*}[!bp]
\begin{swecasebox}[swetaskexample]{Knative Kafka Broker: reconciliation and finalization}
\textcolor{sweviolet!55!black}{\textbf{Problem Statement}}\par\nobreak
\vspace{3pt}
Restore the Kafka Broker control plane so creating, updating, and deleting a Broker produces consistent topic state, endpoint addresses, and readiness conditions. Reconciliation must translate Broker configuration into the shared data-plane contract stored in a ConfigMap. It must preserve other Brokers and existing egress entries, support authentication and transport-security settings, and safely clean up resources during finalization.

\medskip
\textcolor{sweviolet!55!black}{\textbf{Requirements}}\par\nobreak
\begin{minipage}[t]{0.478\linewidth}
\begin{swetaskrequirements}{1}
  \item \textbf{Delivery configuration.} Map content modes and backoff policies to contract values, convert ISO-8601 durations to milliseconds with the expected defaults, and apply egress merge precedence. Contract generation increments must follow the existing wraparound behavior.
  \item \textbf{Topic settings.} Parse \texttt{default.topic.partitions}, \texttt{default.topic.replication.factor}, and \texttt{bootstrap.servers} from ConfigMap data. Reject missing or invalid required values, normalize bootstrap server lists, retain optional \texttt{default.topic.config.*} entries, and provide stable topic names.
  \item \textbf{Topic operations.} Treat an already-existing topic as a successful creation and an unknown topic or partition as a successful deletion. Presence checks must accept valid topic metadata and reject missing, internal, or partitionless topics with useful errors.
  \item \textbf{Addresses and probes.} Produce object paths of the form \texttt{/namespace/name} and service DNS names of the form \texttt{name.namespace.svc}. Build HTTP/HTTPS addressables with the appropriate audience and CA certificate data. Probe adapters must delegate to their supplied functions.
  \item \textbf{Broker reconciliation.} Create or fetch the contract ConfigMap, handling empty or unreadable data. Resolve topic settings, reconcile managed or external topics, and insert or update exactly one Broker entry.
\end{swetaskrequirements}
\end{minipage}\hfill
\begin{minipage}[t]{0.478\linewidth}
\begin{swetaskrequirements}{6}
  \item \textbf{Semantic updates.} Preserve unrelated resources and the Broker's existing egress entries. Increment contract generation only when contract content changes. If receiver or dispatcher pod annotations are stale, update their volume-generation annotations to the persisted generation even when the contract itself is unchanged.
  \item \textbf{Readiness and status.} Set or clear the appropriate conditions and annotations for configuration resolution, data-plane availability, topic readiness, and probe results. Record the resolved dead-letter URI when applicable.
  \item \textbf{Feature settings.} Honor custom topic templates and manage authentication-secret references and finalizers. With OpenID Connect (OIDC) enabled, propagate audiences and apply authorization fallback policies. Project ready event policies, including CESQL filters, into ingress policies. Expose HTTP, HTTPS, or both according to the transport-encryption mode.
  \item \textbf{Finalization.} Remove the Broker entry, persist the updated contract, and synchronize pod generation annotations. Delete managed topics while retaining external topics, and clean up secret finalizers. Tolerate missing ConfigMaps, topics, or secrets and configuration failures specific to teardown. Requeue when probe results or transient Kafka or secret conditions require another attempt.
\end{swetaskrequirements}
\end{minipage}
\end{swecasebox}
\caption{An adapted Go composite task from \texttt{knative-extensions/eventing-kafka-broker}.}
\label{fig:task-example-go}
\end{figure*}

\begin{figure*}[!tp]
\begin{swecasebox}[swetaskexample]{Django REST framework: requests, representations, and versioned URLs}
\textcolor{sweviolet!55!black}{\textbf{Problem Statement}}\par\nobreak
\vspace{3pt}
Restore the runtime behavior that connects incoming Django requests to REST framework parsing, serializer fields, response negotiation, and versioned URLs. Applications must receive consistent values and error details across JSON, form, and multipart inputs. Renderer selection, settings reloads, and hyperlink generation must respect the active request context, including API versions and nested URL namespaces.

\medskip
\textcolor{sweviolet!55!black}{\textbf{Requirements}}\par\nobreak
\begin{minipage}[t]{0.478\linewidth}
\begin{swetaskrequirements}{1}
  \item \textbf{Bound serializer fields.} Support \texttt{serializer[field\_name]} for primitive, JSON, and nested fields. Bound objects must expose values, errors, and fully qualified names, and remain navigable when parent errors are not mappings. Form conversion must turn scalar \texttt{None}/\texttt{False} into empty strings while retaining submitted text for invalid JSON.
  \item \textbf{Request parsing.} Wrap a Django \texttt{HttpRequest}, reject an already wrapped REST framework request, and expose \texttt{query\_params} as \texttt{GET}. Use the selected parser for empty GET/HEAD requests, forms, multipart uploads, and other bodies. Copy parsed form POST/FILES data to the underlying request so repeated form processing is safe. Reprocessing a consumed non-form stream must retain Django's raw-post-data exception.
  \item \textbf{Attribute errors.} Re-raise parser and authenticator attribute errors as \texttt{WrappedAttributeError}, including access through \texttt{user}, \texttt{data}, \texttt{hasattr}, and authentication helpers. Preserve normal attribute proxying and user/auth setters on the underlying request.
  \item \textbf{Error details.} Recursively convert translated strings, lists, tuples, dictionaries, and serializer return containers to \texttt{ErrorDetail}. Preserve message/code structure and consistent equality, hashing, and string representations. Throttling errors must round wait times for messages and expose wait metadata. Bad-request and server-error handlers must return JSON with status 400 and 500.
\end{swetaskrequirements}
\end{minipage}\hfill
\begin{minipage}[t]{0.478\linewidth}
\begin{swetaskrequirements}{5}
  \item \textbf{Content negotiation.} Parse Accept headers and select the best renderer/media-type match, including wildcards and parameters. Format-suffix filtering must raise \texttt{Http404} when no renderer matches. Media-type helpers must preserve matching rules, precedence, and string representations.
  \item \textbf{Settings.} Support string, list, and tuple import coercion through \texttt{perform\_import}, with informative import errors. Reject removed settings, cache resolved values, and refresh local and global \texttt{api\_settings} after a Django settings override. Emit \texttt{rest\_framework.W001} when \texttt{PAGE\_SIZE} is set without \texttt{DEFAULT\_PAGINATION\_CLASS}.
  \item \textbf{Format suffixes.} Generate suffix-aware patterns for \texttt{path()} and \texttt{re\_path()}, preserving order and default keyword arguments. Traverse nested \texttt{include()} trees with mixed path and regex patterns. Support required suffixes and allowed-format lists.
  \item \textbf{Version selection.} Resolve versions from Accept headers, URL paths, namespaces, hostnames, or query parameters. Apply allowed-version and default-version rules, including an explicitly allowed \texttt{None}, and raise the scheme's API exception for invalid versions. Scheme-specific reversing must apply the corresponding version to the URL.
  \item \textbf{Reversing and hyperlinks.} Public \texttt{reverse()} must use the active versioning scheme with ordinary reversing as a fallback and preserve query overrides. Hyperlinked relational fields must resolve namespaced versioned views for serialization and input validation. Non-versioned names must remain reversible regardless of the active version.
\end{swetaskrequirements}
\end{minipage}
\end{swecasebox}
\caption{An adapted Python composite task from \texttt{encode/django-rest-framework}.}
\label{fig:task-example-python}
\end{figure*}

\begin{figure*}[!tp]
\begin{swecasebox}[swetaskexample]{Shepherd: rendering and running interactive tour steps}
\textcolor{sweviolet!55!black}{\textbf{Problem Statement}}\par\nobreak
\vspace{3pt}
Restore Shepherd's step rendering and lifecycle so applications can configure interactive tours, update their contents, and move reliably between steps. A step must coordinate its components, event handlers, attachment target, and positioning with the surrounding Tour. Showing, hiding, or destroying a step must manage accessibility attributes and temporary DOM changes without losing the target's original state.

\medskip
\textcolor{sweviolet!55!black}{\textbf{Requirements}}\par\nobreak
\begin{minipage}[t]{0.478\linewidth}
\begin{swetaskrequirements}{1}
  \item \textbf{Buttons and cancel controls.} Accept literal or callback option values and apply custom \texttt{attrs}, converting numeric and boolean values to strings. Shepherd's own type, class, disabled state, ARIA label, tabindex, and click handler must take precedence over conflicting custom attributes.
  \item \textbf{Content composition.} Render a header, text area, or footer only when its options call for one. Headers may contain a title, cancel icon, or both. Footers contain one button per configured entry and none for an empty or omitted list. Text may be an HTML string, callback result, or \texttt{HTMLElement}.
  \item \textbf{Events.} \texttt{Evented.on} must register handlers, and \texttt{trigger} must forward arguments and preserve their context. \texttt{off} must remove all handlers for an event or a specific handler. A \texttt{once} handler must be removed after its first invocation.
  \item \textbf{Lifecycle.} \texttt{cancel()} and \texttt{complete()} must call the Tour and emit the matching step event. \texttt{hide()} must emit \texttt{before-hide} and \texttt{hide}, hide any modal overlay, and remove target styling. \texttt{destroy()} must emit \texttt{destroy} and remove tooltip and component DOM state.
  \item \textbf{Option updates.} Merge defaults, including nested floating UI options. \texttt{updateStepOptions} must retain unchanged options and refresh rendered text, titles, classes, and buttons when their values change.
\end{swetaskrequirements}
\end{minipage}\hfill
\begin{minipage}[t]{0.478\linewidth}
\begin{swetaskrequirements}{6}
  \item \textbf{Deferred showing.} Await an asynchronous \texttt{beforeShowPromise} before showing the step. Resolve attachment selectors and callbacks when needed for that step, and resolve them again on later shows. Do not evaluate every target eagerly at tour creation. Resolved-attachment lookup must remain safe before the first show.
  \item \textbf{Accessibility and highlights.} Tooltip/dialog markup must include the expected \texttt{aria-labelledby} and \texttt{aria-describedby} links. Apply enabled, target, and highlight classes to the appropriate elements when shown. Hiding or destroying the step must remove these classes and restore prior state.
  \item \textbf{Focus restoration.} Preserve the attachment target's original \texttt{tabindex} across repeated show/hide cycles. If it originally had no attribute, remove the temporary one. If it had a value, restore that value rather than an intermediate runtime value.
  \item \textbf{Advancing.} \texttt{bindAdvance} must support selector-scoped listeners and delegated body listeners. Advance only when the step is open and the configured event/selector conditions match. Unregister listeners when the step is destroyed.
  \item \textbf{Attachment and positioning.} Resolve string selectors and callback targets, binding callbacks to the step. Resolve extra-highlight selector lists to matching elements, and center steps with absent or incomplete attachment options. Preserve an explicit floating UI strategy and place user middleware after Shepherd's default middleware.
\end{swetaskrequirements}
\end{minipage}
\end{swecasebox}
\caption{An adapted JavaScript composite task from \texttt{shipshapecode/shepherd}.}
\label{fig:task-example-js}
\end{figure*}

\begin{figure*}[!tp]
\begin{swecasebox}[swetaskexample]{Element Plus: tab navigation, editing, and reactive updates}
\textcolor{sweviolet!55!black}{\textbf{Problem Statement}}\par\nobreak
\vspace{3pt}
Restore the Element Plus Tabs interaction model and the shared utilities used by tabs and related components. Users must be able to select, add, and close panes through pointer or keyboard input. Active state, event ordering, pane order, and the active indicator must remain consistent when switching is asynchronous or children change reactively. Preserve the public value and configuration conventions used by existing components.

\medskip
\textcolor{sweviolet!55!black}{\textbf{Requirements}}\par\nobreak
\begin{minipage}[t]{0.478\linewidth}
\begin{swetaskrequirements}{1}
  \item \textbf{Active values.} Initialize the active pane from \texttt{modelValue} when defined, otherwise \texttt{defaultValue}, otherwise the first pane index \texttt{'0'}. Support both string and numeric pane names, treating a number and its string representation as distinct values.
  \item \textbf{Switching and guards.} Clicking an enabled tab must emit \texttt{tabClick} immediately. An allowed switch must then emit \texttt{update:modelValue} followed by \texttt{tabChange}. A synchronous \texttt{false} or rejected \texttt{beforeLeave} promise must prevent switching. A pending promise must delay the update/change events until it resolves.
  \item \textbf{Disabled and closable panes.} User interaction must not activate disabled panes. Closing is available only for closable, enabled panes. A pane's explicit \texttt{closable=false} must override the parent's global closable or editable setting.
  \item \textbf{Editing events.} Adding a pane must emit \texttt{edit(undefined, 'add')} and \texttt{tabAdd}. Removing one must emit \texttt{edit(paneName, 'remove')} and \texttt{tabRemove}. Clicking a close button must not also activate the tab.
  \item \textbf{Keyboard and indicator.} Arrow keys must move focus/selection among enabled tabs with wraparound, skipping disabled tabs. Left/up mean previous and right/down mean next. Recalculate the active indicator after DOM updates, using horizontal translation for horizontal tabs and vertical translation for vertical tabs.
\end{swetaskrequirements}
\end{minipage}\hfill
\begin{minipage}[t]{0.478\linewidth}
\begin{swetaskrequirements}{6}
  \item \textbf{Reactive pane order.} Headers must follow the rendered child-pane order after insertion, removal, or reordering, including reactive loops and label-slot updates. These changes must work without remounting the Tabs component.
  \item \textbf{Focus helpers.} \texttt{getFocusIntent} must map supported keys to \texttt{first}, \texttt{last}, \texttt{prev}, or \texttt{next}, respecting orientation and reversing left/right intent for right-to-left layouts. Unsupported or orientation-excluded keys return \texttt{undefined}. \texttt{reorderArray} must rotate an array from the supplied index.
  \item \textbf{Empty-value configuration.} \texttt{useEmptyValues} must expose reactive empty values, the clear value, and an emptiness predicate. Resolve the clear value from props, injected configuration, an explicit default argument, then the library default. Evaluate function-valued clear settings. Emptiness must support primitive inclusion and deep equality for arrays.
  \item \textbf{Shared utilities.} Return typed object entries and support path-based property reads/writes using string or array paths. Global node helpers must create, remove, and retarget appended nodes, including optional IDs. Resolve named/default component sizes from the constants map. Flatten nested vnode structures, including component subtrees and slot/function children.
\end{swetaskrequirements}
\end{minipage}
\end{swecasebox}
\caption{An adapted TypeScript composite task from \texttt{element-plus/element-plus}.}
\label{fig:task-example-ts}
\end{figure*}

\begin{figure*}[!tp]
\begin{swecasebox}[swetaskexample]{Apache Avro: schemas, compatibility, and runtime defaults}
\textcolor{sweviolet!55!black}{\textbf{Problem Statement}}\par\nobreak
\vspace{3pt}
Restore Avro's Java schema APIs so applications can construct schemas, check compatibility, and read data as schemas evolve. Naming, namespaces, properties, and field defaults must agree across the fluent builders and direct creation APIs. Compatibility checks must identify the failing schema node, including in recursive schemas. The same defaults and schema identities must drive record construction, decoding, and single-object encoding.

\medskip
\textcolor{sweviolet!55!black}{\textbf{Requirements}}\par\nobreak
\begin{minipage}[t]{0.478\linewidth}
\begin{swetaskrequirements}{1}
  \item \textbf{Name validation.} \texttt{UTF\_VALIDATOR} must accept names beginning with an underscore or Unicode letter, followed by Unicode letters, digits, or underscores. \texttt{STRICT\_VALIDATOR} restricts letters to ASCII. Both reject null/empty names, leading digits, spaces, and punctuation. Overrides through \texttt{Schema.setNameValidator} must affect only the current thread, including record and field validation.
  \item \textbf{Names and fixed types.} Preserve declared names, documentation, and namespaces. Fully qualified names override the contextual namespace, short names inherit it, and an explicit empty namespace clears it. Short primitive names are reserved only in the empty namespace. Create primitive schemas by type, and reject negative fixed lengths and lengths beyond the existing JVM size limit.
  \item \textbf{Object properties.} Schema and field builders must accept booleans, integral numbers, longs, doubles, strings, byte arrays represented as strings, collections, and maps. Preserve recursive conversion in \texttt{getObjectProp} and reject null property values.
  \item \textbf{Field defaults.} Materialize defaults for primitive, bytes, fixed, enum, map, array, record, and union fields, using the union's first type. Fixed defaults supplied as byte arrays, \texttt{ByteBuffer}, or Avro JSON strings must retain their byte content. Disabling default validation must preserve a raw JSON default even when typed \texttt{defaultVal()} cannot use it.
\end{swetaskrequirements}
\end{minipage}\hfill
\begin{minipage}[t]{0.478\linewidth}
\begin{swetaskrequirements}{5}
  \item \textbf{Compatibility diagnostics.} Accept compatible type promotions and enum-default fallbacks. Report missing reader defaults, fixed-size differences, missing union branches, type mismatches, and nested failures through \texttt{SchemaPairCompatibility}. Preserve incompatibility type, message, reader/writer fragments, description, and the JSON-pointer location of the failing node.
  \item \textbf{Recursive schema histories.} Support can-read, can-be-read, and mutual-read validation strategies with latest-only or full-history checks. Compatible recursive schemas must not overflow the stack. Incompatible histories must raise \texttt{SchemaValidationException}. An unconfigured validation strategy must fail with \texttt{AvroRuntimeException}.
  \item \textbf{Runtime evolution and encoding.} Record builders and datum readers/resolving decoders must supply defaults for added fields, including nested records, maps, unions, enums, bytes, and fixed values. \texttt{BinaryMessageEncoder.encode} must emit Avro single-object encoding with the V1 header, schema parsing fingerprint, and binary payload.
  \item \textbf{Unresolved schemas.} \texttt{getUnresolvedSchemaName} must accept only placeholder records with Avro's expected marker property, placeholder name, and namespace. Ordinary records, fixed schemas, and incorrectly marked placeholders must raise \texttt{IllegalArgumentException}.
\end{swetaskrequirements}
\end{minipage}
\end{swecasebox}
\caption{An adapted Java composite task from \texttt{apache/avro}.}
\label{fig:task-example-java}
\end{figure*}

\begin{figure*}[!tp]
\begin{swecasebox}[swetaskexample]{S3Mock: object request handling and multipart integrity}
\textcolor{sweviolet!55!black}{\textbf{Problem Statement}}\par\nobreak
\vspace{3pt}
Restore S3Mock's object request conversion and multipart upload integrity behavior. The supported AWS SDK request formats must produce consistent access policies, tags, copy sources, and user metadata. Multipart completion must validate upload state, part sizes, order, and ETags with the expected S3 errors. Chunked uploads and multipart digests must expose the correct payload bytes and checksum outputs.

\medskip
\textcolor{sweviolet!55!black}{\textbf{Requirements}}\par\nobreak
\begin{minipage}[t]{0.478\linewidth}
\begin{swetaskrequirements}{1}
  \item \textbf{Canned access policies.} Map \texttt{private}, \texttt{public-read}, \texttt{public-read-write}, \texttt{authenticated-read}, \texttt{aws-exec-read}, \texttt{bucket-owner-read}, and \texttt{bucket-owner-full-control} to their ACL enum values. Unsupported or differently formatted values return \texttt{null}. Policies must give the owner full control and add the required public, authenticated-user, or bucket-owner grants. The \texttt{private} and \texttt{aws-exec-read} policies grant access to the owner alone.
  \item \textbf{Tagging headers.} Accept URL-style tag pairs and XML \texttt{Tagging} payloads. Produce one tag per nonblank pair segment, deserialize XML tag sets, and return \texttt{null} when no tags remain. Malformed XML must raise \texttt{IllegalArgumentException}.
  \item \textbf{Copy sources.} Accept bucket/key values with or without a leading slash, URL-decode them, and extract an optional \texttt{?versionId=} suffix into the version field. Null input must raise \texttt{NullPointerException}. Inputs that cannot be parsed into bucket and key must raise \texttt{IllegalArgumentException}.
  \item \textbf{Part numbers.} \texttt{verifyPartNumberLimits} must accept and return integer part numbers from 1 through 10,000. Values outside this range must raise \texttt{INVALID\_PART\_NUMBER}.
  \item \textbf{Upload lookup.} \texttt{verifyMultipartUploadExists} must validate an upload in its bucket using the requested \texttt{includeCompleted} option and return the upload information. Translate a missing upload in the store to \texttt{NO\_SUCH\_UPLOAD\_MULTIPART}.
\end{swetaskrequirements}
\end{minipage}\hfill
\begin{minipage}[t]{0.478\linewidth}
\begin{swetaskrequirements}{6}
  \item \textbf{Stored-part validation.} Check upload existence before validating completion. Every part except the last must meet \texttt{MINIMUM\_PART\_SIZE} and the required part-number bounds. Size and number violations must raise \texttt{ENTITY\_TOO\_SMALL} and \texttt{INVALID\_PART\_NUMBER}, respectively.
  \item \textbf{Requested completion parts.} Resolve the upload target from the object key. An unmapped target, missing part, or mismatched part number/ETag must raise \texttt{INVALID\_PART}. Match requested parts to stored parts and require ascending order, raising \texttt{INVALID\_PART\_ORDER} when order is invalid.
  \item \textbf{Chunked payloads.} Decode AWS unsigned chunked streams into raw bytes while tracking expected decoded length, bytes read, and chunk count. At the terminal zero-size chunk, parse an optional \texttt{x-amz-checksum-} trailer into algorithm and checksum fields. Leave those fields unset when no checksum trailer is present.
  \item \textbf{Multipart digests.} Hash each part independently, concatenate the part digests, and hash that concatenation using the requested algorithm. Append the part-count suffix to the final digest. Return hexadecimal output for the MD5 ETag variant and Base64 output for the checksum variant.
  \item \textbf{User metadata.} Read \texttt{x-amz-meta-} headers case-insensitively, including canonical and lowercase forms, and ignore blank values. When emitting metadata headers, retain the prefix if present and add it when absent so metadata can round-trip.
\end{swetaskrequirements}
\end{minipage}
\end{swecasebox}
\caption{An adapted Kotlin composite task from \texttt{adobe/S3Mock}.}
\label{fig:task-example-kotlin}
\end{figure*}

\end{document}